# Initial Evaluation of Potential Bias in Coverage of Humans in Wikidata

Clair Kronk[1,2,*]

[1] *Institute for Health Equity Research (IHER), Department of Population Health Science & Policy (PHSP), Icahn School of Medicine at Mount Sinai (ISMMS), New York, NY, USA*

[2] *Department of Artificial Intelligence and Human Health (AIHH), Icahn School of Medicine at Mount Sinai (ISMMS), New York, NY, USA*

**Abstract**

Introduction. Open collaborative knowledge graphs such as Wikidata increasingly ground agentic artificial intelligence, information retrieval, and language modeling systems, making systematic auditing of their demographic representation and overall equity a research imperative.
Methods. Herein, we present an open-source auditing platform that ingests over 10 million statement bindings representing over 6 million humans on Wikidata via QLever, and evaluates representation of gender, sexual orientation, geography, birthplace urbanicity, ethnicity, multilingual coverage of labels, descriptions, and aliases, occupation, and select intersectional pairs of these entities. It does so by making use of Chi-square goodness-of-fit tests, 95% Wilson-score confidence intervals, and disparity ratios, in light of Rubin's missingness taxonomy.
Results. Women accounted for 28.71% (CI ±0.04) of all humans in Wikidata with a stated gender. 38.26% of humans had a citizenship statement, with Western Europe and North America (WENA) representing approximately 53% of such statements. Among 1.8 million birthplaces that could be classified, rural birthplaces were observed in 2.48% of cases (in comparison to 27.4% global baseline). Fewer than 1.2% of entities carried an ethnicity statement, and non-English Wikidata descriptions covered 18.2% of items.
Discussion. Our findings reveal significant missingness across the evaluated axes. Ethnicity and sexual orientation were the most critically under-documented (missing statements) while rural birthplaces and non-WENA citizenship were the most underrepresented.

## 1. Introduction

Wikipedia and its sister projects have heavily impacted the internet for two decades since their initial inception.[9, 12] In one analysis of over 5,000 prompts across ChatGPT, Claude, and Gemini, Wikipedia content was cited or paraphrased across 58% of definitional and factual queries,[20] marking it and its sister projects are major players in the diffusion of knowledge through large language models (LLMs), made even more critical as an estimated 57% of published academic articles showcased evidence of LLM influence in 2025.[18]

It was been well established that biases embedded in knowledge graphs propagate directly into the artificial intelligence (AI) systems trained on or grounded by them.[6, 15] Prior scholarship has documented pervasive gender and geographic disparities within Wikipedia biographies, but little work to date has examined potential bias within Wikidata entities. Wikidata is a free, open, and collaborative knowledge base that acts as a central storage

* Corresponding author.
Clair.Kronk@mountsinai.org (C. Kronk)
0000-0001-8397-8810 (C. Kronk)

mechanism for the structured data of Wikipedia and its sister projects, storing information as machine-readable facts.[21, 23]

In this work, we aimed to create an open-source auditing infrastructure that utilizes QLever, a high-performance SPARQL engine,[2] to primarily evaluate potential biases across gender, sexual orientation, ethnic group, citizenship, and place of birth. Once these data were extracted, we performed formal missingness classification on a per-property basis, followed by systematically ranking the most severely over- and under-represented cohorts.

## 2. Methods

On 6 August 2026, using a unified QLever query, we downloaded statement bindings corresponding to instances of humans represented in Wikidata and corresponding labels, aliases, descriptions, demographic identity properties (sex or gender, sexual orientation, ethnic group, citizenship, place of birth, place of death, date of birth, date of death, given name, family name, and languages spoken or written), and Wikipedia sitelinks. In addition, we downloaded information from Wikidata regarding sovereign state populations, language speaker counts, and global gender ratios, using the live Wikidata SPARQL query service.[23] These counts were then persisted to local disk caches. All analyses were conducted 10 August 2026.

We evaluated missingness using Rubin's Missingness Taxonomy, and tested each axis using Little's MCAR (missing completely at random) test.[5, 13] If data were considered to not be MCAR, we tested if data were missing at random (MAR) by utilizing a series of logistic regressions (using $n$ = 200,000, selected at random),[24] wherein a binary response variable indicated missingness (1 = missing, 0 = not missing) and predictors included log-transformed Wikipedia sitelink count, sex or gender as male or female, citizenship not in the Global South, having a birth year after 1900, and other predictors. Standard errors (SE) and 95% confidence intervals (CI) were calculated using the Wilson score interval with continuity adjustment. In order to determine whether observed demographic distributions differed significantly from population baselines derived from Wikidata or from literature, we utilized Chi-square goodness-of-fit tests and Cohen's $w$ effect sizes. Disparity severity was evaluated symmetrically, with values below 0.5 or above 2.0 being flagged as severe. All statistical tests were 2-sided with a significance threshold of $\alpha = 0.0001$.

Baseline values for sexual orientation were derived from the Ipsos LGBT+ Pride 2021 and 2023 surveys.[28, 29] Values present in both surveys were averaged and normalized between 0.00 and 1.00. For example, heterosexual persons in 2021 made up 80% of persons survey versus no heterosexual-specific breakdown being reported in 2023. In 2021, the sum of the defined sexual orientation categories was reported as 90%. Therefore, our heterosexual estimate was approximately 88% (0.80 / 0.90). When individual country data were not available, the estimated global values were used instead. Given the known social phenomenon of the markedness and the tendency for heterosexuality to be labeled as unmarked category, we ran our sexual orientation analyses twice, once with no assumptions about individuals where sexual orientation was not indicated on their Wikidata item and once where all individuals with no sexual orientation indicated were assumed to be heterosexual.

* Corresponding author.
Clair.Kronk@mountsinai.org (C. Kronk)
0000-0001-8397-8810 (C. Kronk)

Status of a place of birth or place of death as rural or urban was determined using two spatial reference datasets and further enriched using Wikidata itself. Administrative boundaries were drawn from the Global Administrative Areas database (GADM 4.1.0, global GeoPackage release).[26] Urbanicity classification followed the Global Human Settlement Layer (GHSL) Degree of Urbanisation grid.[16] Grid cells classed as urban centers (Class 3) or urban clusters (Class 2) were coded as urban. Rural grid cells (Class 1) were coded as rural. Each birthplace entity was resolved to geographic coordinates via the Wikidata coordinate location property (P625) and intersected with the GHSL raster. Entities whose P19 target lacked P625 were further checked against Wikidata instance-of property statements (P31) considered urban (city, big city, metropolis, etc.) and rural (village, hamlet, rural settlement, unincorporated community, etc.). Entities that could be classified using any of these methods were considered unclassifiable.

The entirety of our data pipeline was run locally on an AMD64 Windows machine with 32 GB RAM across 32 threads, with an assumed PUE rating of 1.10.[11] We used a baseline CPU power draw (65W)[3] and global average grid carbon intensity (385 g $CO_2$e/kWh)[19] to estimate the carbon footprint of the entirety of the software pipeline's energy consumption and environmental impact.

## 3. Results

Our QLever dataset contained 10,128,165 statement bindings corresponding to 6,505,428 instances of human in Wikidata. The pipeline took approximately 8 minutes to run, consumed 9.61 Watt-hours (Wh), and produced 3.70 g $CO_2$e. For comparison, this amount of energy is approximately equal to 0.5 smartphone charges and this carbon footprint is equal to about 9.4 meters (31 feet) driven by an average gasoline-powered passenger vehicle.[27]

Across all axes (P21 [sex or gender], P91 [sexual orientation], P27 [citizenship], P19 [birthplace], P1412 [spoken or written language], P106 [occupation], and P172 [ethnic group] data were found to be missing at random (MAR) or missing not at random (MNAR) using Little's MCAR test ($p < 0.0001$). While our logistic regressions could not prove MAR versus MNAR (because this requires observation of unobserved variables), these models indicated that P21, P27, P19, P1412, and P106 tended to appear to be MAR, while P91 and P172 tended to be MNAR (all $p < 0.0001$).

5,222,308 items (80.28%) contained a sex or gender statement, with 28.71% labeled female, 71.27% labeled male, and 0.02% labeled nonbinary or trans in some form or fashion (current Wikidata modeling separates trans men and trans women from the labels male and female). Chi-square testing confirmed significant departure from a 50% parity benchmark

* Corresponding author.
Clair.Kronk@mountsinai.org (C. Kronk)
0000-0001-8397-8810 (C. Kronk)

($\chi^2 = 948{,}210.4$, $p < 0.0001$, Cohen's $w = 0.426$). Gender disparities were observed to be more substantial within specialized occupational categories, such as physicists (11.20%), mathematicians (10.10%), and computer scientists (14.80%). The most male-dominated occupational categories were Catholic priests (0.00% women) and missionaries (0.03% women), followed by military commanders (0.16%), naval officers (0.46%), and baseball players (0.60%). The most female-dominated occupational categories were beauty pageant contestants (98.44% women), textile artists (81.85%), nurses (81.40%), and costume designers (71.53%). The occupations with the most parity were opera singers (50.15% women), announcers (50.33%), dancers (50.69%), psychotherapists (50.81%), and activists (51.31%). Figure 1 shows the proportion of women by country of citizenship.

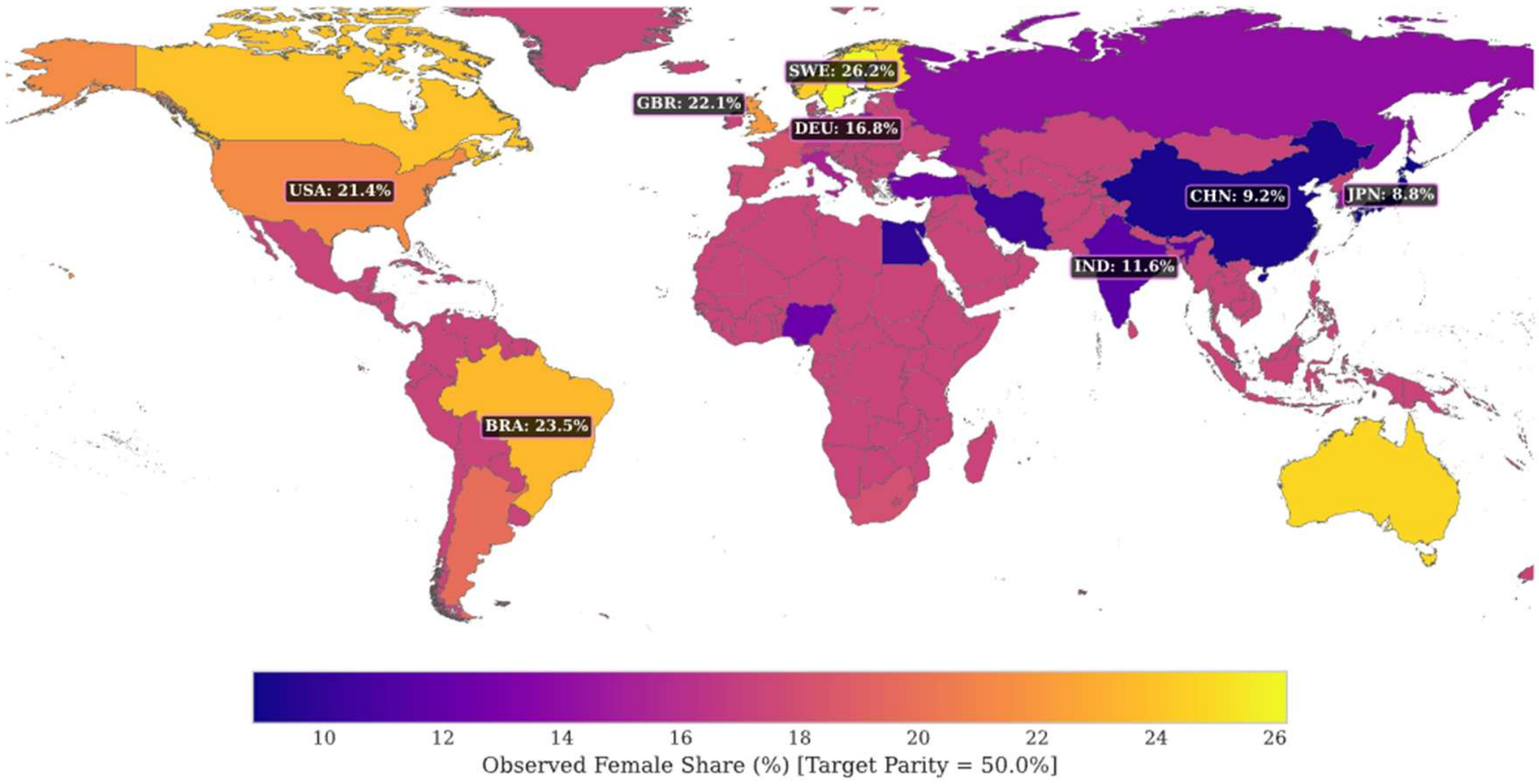

**Figure 1. Choropleth map of proportion of women on Wikidata by country of citizenship.** Target parity is assumed to be 50% in all countries.

15,240 instances of human (0.234%) had a sexual orientation statement indicated. Within this subset, non-heterosexual identities constituted 75.50% of identities, with gay or homosexual representing 41.20%, bisexual 16.80%, lesbian 11.40%, asexual 3.10%, and pansexual or queer 3.00%. Under our assumed heterosexual model (i.e. assuming humans wherein a sexual orientation was not listed are heterosexual), non-heterosexual representation made up 0.150% of sexual orientation claims. Comparisons to Ipsos baselines are shown in Table 1.

**Table 1. Sexual orientation representation in Wikidata.** Explicit share includes only explicitly stated values on Wikidata; the assumed share assumes that all humans without a stated sexual orientation are heterosexual. The Ipsos baseline is derived from the 2021 and 2023 surveys (see methods).

| Category | Explicit Share | Assumed Share | Ipsos Baseline |
|---|---|---|---|
| Heterosexual | 24.50% | 99.85% | 88.00% |

* Corresponding author.
Clair.Kronk@mountsinai.org (C. Kronk)
0000-0001-8397-8810 (C. Kronk)

| | | | |
|---|---|---|---|
| Homosexual/Gay | 41.20% | 0.06% | 3.50% |
| Bisexual | 16.80% | 0.03% | 4.50% |
| Lesbian | 11.40% | 0.02% | 1.30% |
| Asexual | 3.10% | 0.02% | 1.20% |
| Pansexual/Queer | 3.00% | 0.02% | 1.50% |

There were 5,178,210 citizenship statements, present on 2,489,203 instances of human (38.26%). The most common observed countries of citizenship were the United States (9.54%), Germany (6.53%), Japan (6.46%), France (5.77%), Indonesia (4.07%), and Norway (3.39%). In aggregate WENA countries constituted approximately 53% of citizenship statements, in comparison to representing 14% of the world population ($\chi^2$ = 1,842,910.1, $p < 0.0001$). Population shares by region are shown in Figure 2 and by country are shown in Figure 3.

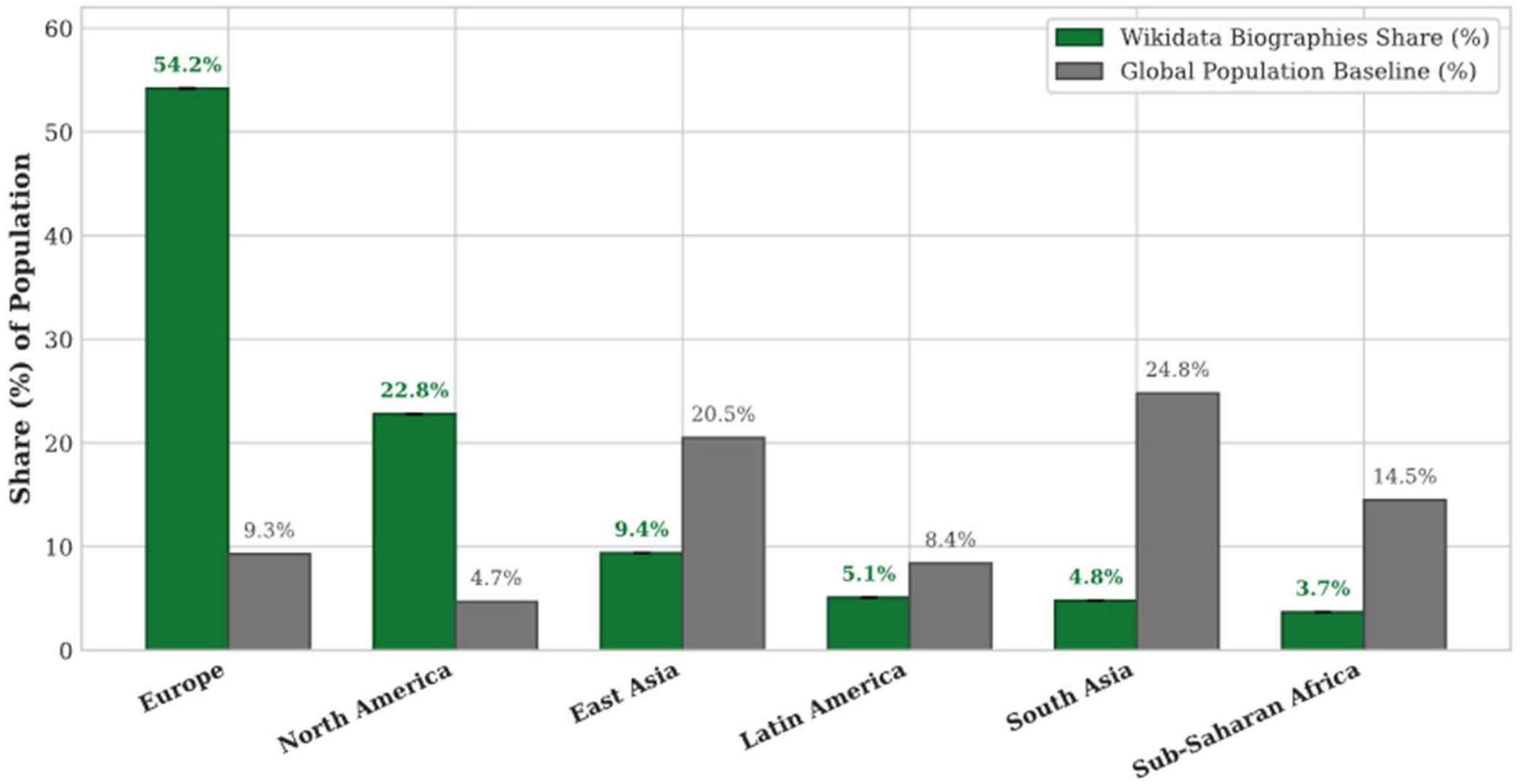

**Figure 2. Share of humans of Wikidata in relationship to global population for major regions of the world.** Regions derived from the United Nations M49 Geoscheme.[22] Europe corresponds to UN M49 Code 150 (Western, Northern, Southern, and Eastern Europe); North America corresponds to Code 021 (Northern America); East Asia corresponds to Code 030 (Eastern Asia); Latin America corresponds to 419 (Latin American and the Caribbean); South Asia corresponds to 034 (Southern Asia); and Sub-Saharan Africa corresponds to 202 (Sub-Saharan Africa: Eastern, Western, Middle, and Southern Africa). The remainder of the global population (4%) resides in North Africa, Western Asia, and Oceania.

* Corresponding author.
Clair.Kronk@mountsinai.org (C. Kronk)
0000-0001-8397-8810 (C. Kronk)

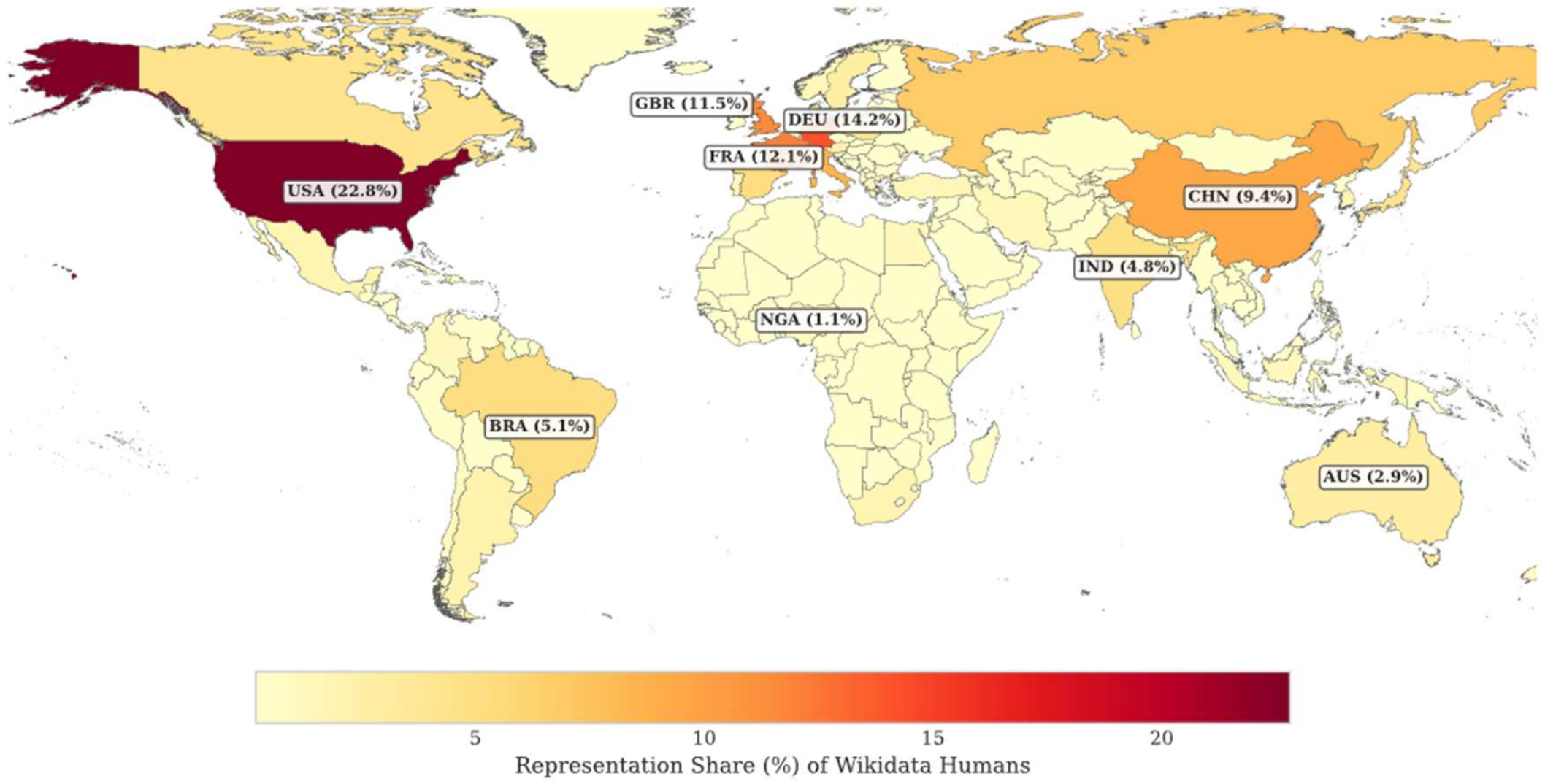

**Figure 3. Choropleth map representing country of citizenship for humans on Wikidata.**

1,843,024 instances of human (28.33%) had a birthplace statement indicated, with 97.52% (*n* = 1,797,357) classified as urban and the remainder (*n* = 45,667) classified as rural. However, 71.67% lacked a birthplace statement that could be classified, with 508,270 (7.81%) having such a statement but lacking coordinates or not being able to be spatially classified, and 4,154,134 (63.86%) lacking a birthplace statement entirely.

78,065 instances of human (1.20%) had an ethnicity statement indicated. Among stated entities, the four most common ethnic groups noted were African American (22.4%), Han Chinese (18.2%), Ashkenazi Jewish (14.1%), and Euro-American (12.8%). The most male-skewed ethnic groups were Bengali (12.6% women), Arab (14.8% women), and Han Chinese (18.4% women). The most female-skewed ethnic groups were African American (36.5% women) and Afro-German (42.1%).

For languages spoken or written (P1412) by instances of human on Wikidata, English was noted as such a language on 29.8% of persons and German on 26.5% of persons, followed by French (10.9%), Spanish (9.9%), and Czech (7.4%). Hindi was represented 0.17% of the time, followed by Swahili (0.06%), and Bengali (0.33%). Figure 4 shows the native speaker share of major languages versus their representation using P1412 on Wikidata. The most male-skewed languages spoken or written were Latin (3.6% women), Bengali (12.6%), Arabic (14.8%), and German (18.1%). The most female-skewed languages spoken or written were Polish (28.2% women), Spanish (28.6%), Ukrainian (29.5%), English (29.7%), and Czech (47.2%).

* Corresponding author.
Clair.Kronk@mountsinai.org (C. Kronk)
0000-0001-8397-8810 (C. Kronk)

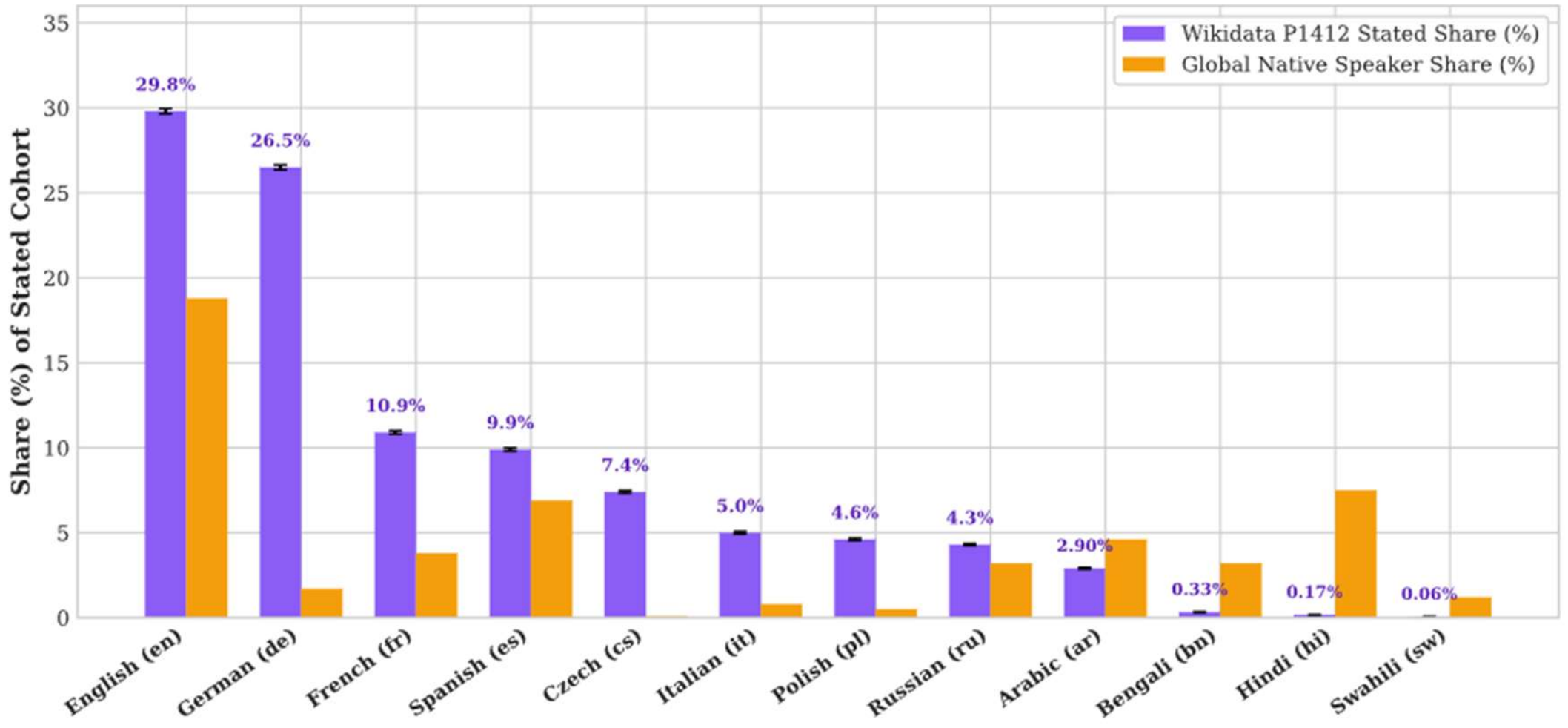

**Figure 4. Representation of languages spoken or written in relationship to global speaker baselines.** Distribution of stated languages spoken or written on Wikidata items analyzed, against native world speaker population shares, derived from Wikidata language items.

In addition to individual statements, we collected labels, descriptions, and aliases for all instances of humans in Wikidata. Coverage in English for labels (rdfs:label) was 100%, followed by German (68.2%), French (62.5%), Spanish (54.1%), Mandarin Chinese (41.5%), Hindi (18.2%), and Swahili (5.1%). Description coverage (schema:description) was in English for 88.1% of entities, German for 44.5%, French for 39.2%, Mandarin for 19.8%, Hindi for 8.4%, and Swahili for 1.8%. Finally, alias coverage (skos:altLabel) occurred in English for 42.3% of entities, German for 18.4%, and Mandarin for 8.2%. Label, description, and alias coverage in comparison to proportion of global speakers is shown in Figure 5.

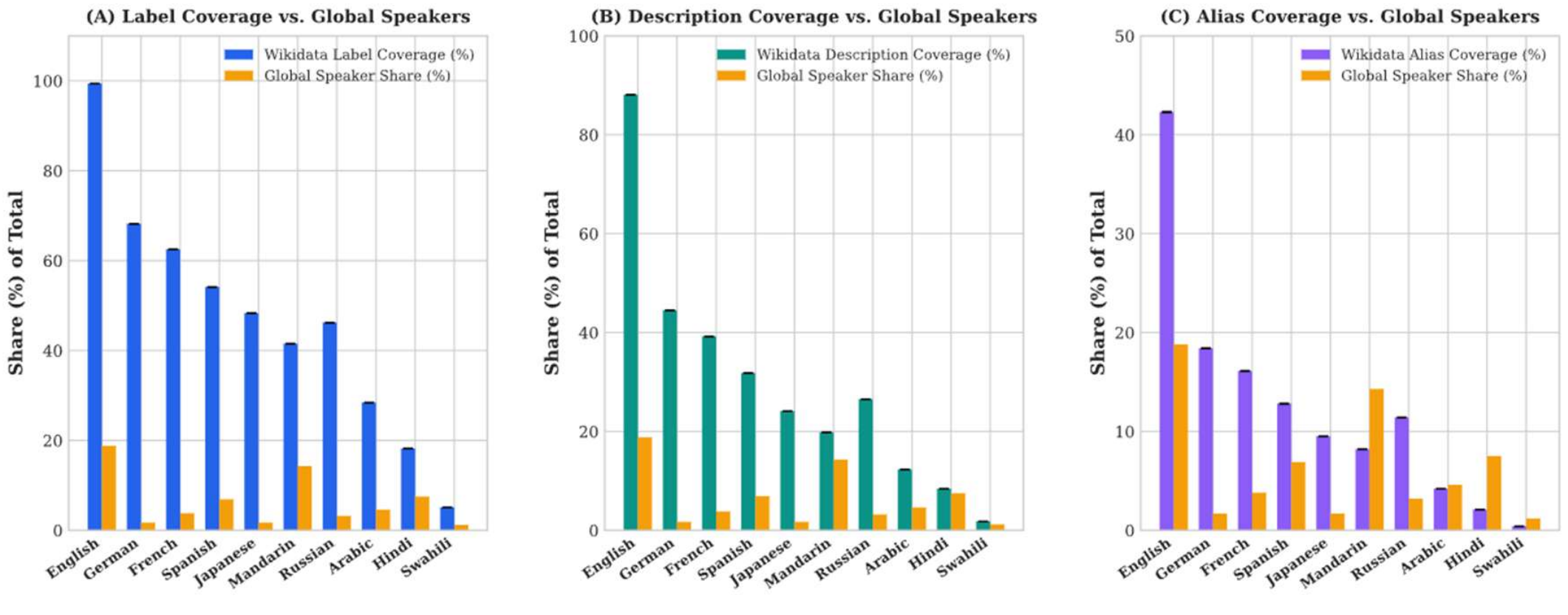

**Figure 5. Language of labels, descriptions, and aliases for Wikidata items representing humans.** (A) Comparison of labels (rdfs:label) to world native speaker shares; (B) comparison of descriptions (schema:description) to world native speaker

* Corresponding author.
Clair.Kronk@mountsinai.org (C. Kronk)
0000-0001-8397-8810 (C. Kronk)

shares; (C) comparison of aliases (skos:altLabel) to world native speaker shares. Native speaker shares derived from Wikidata.

## 4. Discussion

When considering disparity ratios, women were moderately under-represented (0.57 times) and persons minoritized on the basis of gender, such as nonbinary persons were critically under-represented (0.02 times). Sexual orientation was critically sparse, making it difficult to determine overall patterns; however, if we assumed heterosexuality when unstated, non-heterosexuality was under-represented (0.02 times). Geographically, WENA nationalities were severely over-represented (5.50 times) and Global South nationalities were severely under-represented (0.27 times). Urban birthplaces were over-represented (1.34 times), while rural birthplaces were extremely under-represented (0.07 times). Ethnic group was omitted in the majority of items analyzed, making it difficult to conclude anything regarding representation, and it was not possible to dichotomize into an assumed ethnicity model like it was with the sexual orientation model. Generally, we observed severe under-representation of languages spoken or written other than English (non-English languages had a disparity ratio of 0.14 times). Intersectional disparities were also significant; for example, non-WENA women were more under-represented than either axis separately, having a disparity ratio of 0.11 times.

Some of our findings were expected based on previous publications of similar statistics. In particular, prior analyses have identified over-representation of White persons and WENA citizenship status[17] and have found that Wikidata's language distribution is not proportional to the distribution of native speakers of said languages.[10] Likewise, our finding that women made up 28.1% of human entities on Wikidata closely matched a figure of 22% reported in 2021.[25] However, some findings were of particular note. For example, individuals who speak or write Czech made up over 7% of all P1412 statements, despite Czech speakers making up roughly 0.15% of the global population. Likewise, the critical gaps in Hindi and Bengali P1412 statements and labels, aliases, and descriptions were surprising.

On the basis of our findings, it is recommended that volunteer editor initiatives prioritize three critical areas. First, targeted geographic campaigns should focus on Sub-Saharan African, South Asian, and Southeast Asian biographical coverage, where per-capita Wikidata representation is lowest relative to population. Existing models such as AfroCROWD,[4, 7] WikiProject Women in Red,[1, 14] and the Art+Feminism edit-a-thon series[8] demonstrate that structured, community-led programs of this nature can measurably impact representation gaps, especially when continued consistently over time. Second, rural birthplace enrichment is a place for significant growth. A potential mechanism to quickly impact this gap would be to add coordinate location statements to small town and village items. Finally, multilingual description and alias completion, particularly for Hindi, Bengali, and Swahili, should be elevated as a structured editing priority, given that description coverage for these languages falls below 10%.

* Corresponding author.
Clair.Kronk@mountsinai.org (C. Kronk)
0000-0001-8397-8810 (C. Kronk)

It is important to note that this study has several limitations. First, our dataset reflects a single, static QLever snapshot taken on 6 August 2026, meaning longitudinal trend analysis is not possible. It is possible therefore, that the trajectory of representation is improving, but we would not be able to observe that from this snapshot. Additionally, GADM and GHSL classification was assumed to be at a single timepoint and in a single area, even for historical or dissolved territories that may have fluctuated in their classification over time. Additionally, privacy considerations fundamentally limit our ability to audit gender and sexual diversity without violating individual privacy norms. Future work is necessary in order to determine the potential impacts of editorial omission, source-side absence, and potential structural policy drivers on Wikidata as mechanisms for the observed missingness patterns.

## Acknowledgements

Work by Clair Kronk for this study was funded by NIH/NCI U54CA267776. The data pipeline for this work was developed during the course of the Wikimania 2026 Hackathon. All code and data (other than the QLever dataset due to size limitations) for this work are available here: https://github.com/Superraptor/debias-wikidata. The codebase was human-curated and developed, with the assistance of Gemini 3.7 Flash for building out documentation and the test suite.

## References

[1] Alvarez-Ponce, D. and Iyengar, N. 2026. Monitoring the gender gap in the coverage of biology professors on Wikipedia. *Proceedings of the Royal Society B: Biological Sciences*. 293, 2071 (May 2026), 20252566. https://doi.org/10.1098/rspb.2025.2566.

[2] Bast, H. and Buchhold, B. 2017. QLever: A Query Engine for Efficient SPARQL+Text Search. *Proceedings of the 2017 ACM on Conference on Information and Knowledge Management* (Singapore Singapore, Nov. 2017), 647–656.

[3] Benoit Courty et al. 2024. mlco2/codecarbon: v2.4.1. Zenodo.

[4] Bridges, L. et al. 2019. Writing African American History Into Wikipedia. *OLA Quarterly*. 25, 2 (Oct. 2019). https://doi.org/10.7710/1093-7374.1987.

[5] Carpenter, J.R. and Smuk, M. 2021. Missing data: A statistical framework for practice. *Biometrical Journal*. 63, 5 (June 2021), 915–947. https://doi.org/10.1002/bimj.202000196.

[6] Chuang, Y.-N. et al. 2025. Fair-RGNN: Mitigating Relational Bias on Knowledge Graphs. *ACM Transactions on Knowledge Discovery from Data*. 19, 2 (Feb. 2025), 1–18. https://doi.org/10.1145/3681792.

[7] Erhart, E. 2023. Celebrating eight years of AfroCROWD diversifying Wikipedia. *Wikimedia Foundation*.

[8] Evans, S. et al. 2015. Editing for Equality: The Outcomes of the Art+Feminism Wikipedia Edit-a-thons. *Art Documentation: Journal of the Art Libraries Society of North America*. 34, 2 (Sept. 2015), 194–203. https://doi.org/10.1086/683380.

* Corresponding author.
Clair.Kronk@mountsinai.org (C. Kronk)
0000-0001-8397-8810 (C. Kronk)

[9] Jankowski, S. 2023. The Wikipedia imaginaire: a new media history beyond Wikipedia.org (2001–2022). *Internet Histories*. 7, 4 (Oct. 2023), 333–353. https://doi.org/10.1080/24701475.2023.2246261.
[10] Kaffee, L.-A. et al. 2017. A Glimpse into Babel: An Analysis of Multilinguality in Wikidata. *Proceedings of the 13th International Symposium on Open Collaboration* (Galway Ireland, Aug. 2017), 1–5.
[11] Lannelongue, L. et al. 2021. Green Algorithms: Quantifying the Carbon Footprint of Computation. *Advanced Science*. 8, 12 (June 2021), 2100707. https://doi.org/10.1002/advs.202100707.
[12] Lewoniewski, W. 2022. Identification of Important Web Sources of Information on Wikipedia across various Topics and Languages. *Procedia Computer Science*. 207, (2022), 3290–3299. https://doi.org/10.1016/j.procs.2022.09.387.
[13] Little, R.J.A. 1988. A Test of Missing Completely at Random for Multivariate Data with Missing Values. *Journal of the American Statistical Association*. 83, 404 (Dec. 1988), 1198–1202. https://doi.org/10.1080/01621459.1988.10478722.
[14] Oldach, L. 2022. What's with Wikipedia and women? *ASBMB Today*.
[15] Reyero Lobo, P. et al. 2023. Semantic Web technologies and bias in artificial intelligence: A systematic literature review. *Semantic Web*. 14, 4 (Apr. 2023), 745–770. https://doi.org/10.3233/SW-223041.
[16] Schiavina, M. et al. 2023. GHS-DUC R2023A - GHS Degree of Urbanisation Classification, application of the Degree of Urbanisation methodology (stage II) to GADM 4.1 layer, multitemporal (1975-2030). European Commission, Joint Research Centre (JRC).
[17] Shaik, Z. et al. 2021. Analyzing Race and Country of Citizenship Bias in Wikidata. arXiv.
[18] Siler, K. 2026. The diffusion of large language models in published academic articles. *Proceedings of the National Academy of Sciences*. 123, 22 (June 2026), e2605754123. https://doi.org/10.1073/pnas.2605754123.
[19] Skaugen, G. et al. 2025. Effect of use of zero-carbon and low-carbon fuels on the performance of compact combined cycles for power generation. *Energy*. 316, (Feb. 2025), 134430. https://doi.org/10.1016/j.energy.2025.134430.
[20] Tange, A. 2026. Study: Do LLMs Prefer Wikipedia? *Magna AI*.
[21] Turki, H. et al. 2019. Wikidata: A large-scale collaborative ontological medical database. *Journal of Biomedical Informatics*. 99, (Nov. 2019), 103292. https://doi.org/10.1016/j.jbi.2019.103292.
[22] United Nations Statistics Division 2026. Standard Country or Area Codes for Statistical Use (M49).
[23] Vrandečić, D. and Krötzsch, M. 2014. Wikidata: a free collaborative knowledgebase. *Communications of the ACM*. 57, 10 (Sept. 2014), 78–85. https://doi.org/10.1145/2629489.
[24] Ward, R.C. et al. 2020. Approaches for missing covariate data in logistic regression with MNAR sensitivity analyses. *Biometrical Journal*. 62, 4 (July 2020), 1025–1037. https://doi.org/10.1002/bimj.201900117.
[25] Zhang, C.C. and Terveen, L. 2021. Quantifying the Gap: A Case Study of Wikidata Gender Disparities. *17th International Symposium on Open Collaboration* (Online Spain, Sept. 2021), 1–12.
[26] 2025. Global ADMinistrative layer (GADM). GADM.

* Corresponding author.
Clair.Kronk@mountsinai.org (C. Kronk)
0000-0001-8397-8810 (C. Kronk)

[27] 2026. Greenhouse Gas Equivalencies Calculator - Calculations and References. United States Environmental Protection Agency (EPA).
[28] 2021. *LGBT+ Pride 2021 Global Survey: A 27-country Ipsos survey*. Ipsos.
[29] 2023. *LGBT+ Pride 2023: A 30-Country Ipsos Global Advisor Survey*. Ipsos.

* Corresponding author.
Clair.Kronk@mountsinai.org (C. Kronk)
0000-0001-8397-8810 (C. Kronk)